\documentclass[12pt,preprint,a4paper]{aastex}
\usepackage{graphics,epsf}
\usepackage{amsmath}                
\usepackage{amsfonts}               
\usepackage{amssymb}                
\usepackage{epsfig}                 
\usepackage{subfigure}

\def \kms{$~\rm{km}~\rm{s}^{-1}$}

\def \cmcub {$~\rm{cm}^{-3}$}

\def \cm{~\rm{cm}}
\def \s{~\rm{s}}
\def \km{~\rm{km}}
\def \kms{$~\rm{km}~{\rm s}^{-1}$}

\def \K{~\rm{K}}
\def \g{~\rm{g}}

\def \AU{~\rm{AU}}

\def \yr{~\rm{yr}}

\def \etc{$\eta$~Car}

\begin{document}

\title{NUMERICAL SIMULATIONS OF WIND--EQUATORIAL GAS INTERACTION IN ETA CARINAE}

\author{Danny Tsebrenko\altaffilmark{1}, Muhammad Akashi\altaffilmark{1}, and  Noam Soker\altaffilmark{1}}

\altaffiltext{1}{Department of Physics, Technion -- Israel Institute of Technology, Haifa 32000, Israel;
ddtt@tx.technion.ac.il; akashi@ph.technion.ac.il; soker@physics.technion.ac.il.}

\begin{abstract}
We perform three-dimensional gas-dynamical simulations and show that the asymmetric morphology of the blue and red-shifted components of the outflow
at hundreds of astronomical units (AU) from the massive binary system $\eta$ Carinae can be accounted for from the collision of the free
primary stellar wind with the slowly expanding dense equatorial gas.
Owing to the very complicated structure of the century-old equatorial ejecta, that is not fully spatially resolved by observations,
we limit ourselves to modelling the equatorial dense gas by one or two dense spherical clouds.
Because of that we reproduce the general qualitative properties of the velocity maps, but not the fine details.
The fine details of the velocity maps can be matched by simply structuring the dense ejecta in an appropriate way.
The blue and red-shifted components are formed in the post-shock flow of the primary wind, on the
two sides of the equatorial plane, respectively.
The fast wind from the secondary star plays no role in our model, as for most of the orbital period in our
model the primary star is closer to us. The dense clouds are observed to be closer to us than the binary system is,
and so in our model the primary star faces the dense equatorial ejecta for the majority of the orbital period.
\end{abstract}

\keywords{stars: mass loss --- stars: winds, outflows --- stars:
variables: other --- stars: individual (Eta Car)}

\section{INTRODUCTION}
\label{sec:intro}

A picture is emerging in recent years according to which binary interaction plays a major role in the evolution of very massive stars.
The main processes involved are mass transfer and merger  (e.g., \citealt{KashiSoker2010a, SokerKashi2012a, Sana2012}).
The best example is $\eta$ Carinae, one of the most massive binary systems in our galaxy, composed of a primary
Luminous Blue Variable (LBV), and a somewhat evolved massive O-star secondary.
The estimated masses are $M_1 \simeq 120-200 M_\odot$ and $M_2 \simeq 60-90 M_\odot$ \citep{KashiSoker2010a}, respectively.
The orbital period is five and a half years \citep{Damineli2008}, and the orbit is highly eccentric, $e \simeq 0.9$.
For several weeks every periastron passage the binary system experiences a strong interaction, possibly with some mass transfer.
Along the rest of the orbit the winds from the two stars collide with each other. The colliding winds determine many of the
emission and absorption properties of \etc.
The X-ray emission, for example,  comes directly from the post-shock regions of the two winds
(\citealt{Corcoran2005, Corcoran2010, Parkin2011, Akashi2006, Akashi2012, Moffat2009, Henley2008, Okazaki2008, Pittard2002, Pittard1998, Behar2007}).

The binary interaction was much stronger during the two eruptions in the nineteenth century, when the primary is thought to
have been in an unstable state. Huge amounts of mass were lost during the 1837.9 -- 1858 Great Eruption
(GE; see \citealt{DavidsonHumphreys1997} for a review on the GE and some other observed properties of \etc).
The bipolar Homunculus nebula surrounding the system and containing most the ejected mass is the most prominent structural feature from the GE
(\citealt{Gomez2006, Gomez2010, SmithOwocki2006, Smith2003, SmithFerland2007}).
Some non-negligible fraction of the mass lost by the primary star was accreted by the secondary star during these twenty years \citep{KashiSoker2010a}.
The accretion rate was higher near periastron passages, accounting for observed luminosity peaks during the GE.

Both the GE and the weaker 1887.3 -- 1895.3 Lesser Eruption (LE) that followed it are the source of most of the circumbinary medium (CBM).
The LE was a much less energetic eruption (\citealt{Humphreys1999}) and only $0.1$ -- $1 \rm{M_\odot}$ were ejected from the primary (\citealt{Smith2005}).
Material ejected during the LE is thought to be the source of the dense inhomogeneous blobs and the gas in their surroundings closer to the binary system (\citealt{Smith2004}).
These blobs are known as the Weigelt blobs (WBs; \citealt{WeigeltEbersberger1986, HofmannWeigelt1988}).
The WBs are concentrated on the NW side of $\eta$ Car that is closer to us, but there is also dense material in other directions
(e.g., \citealt{Smith2000, Dorland2004, Chesneau2005, Gull2009, Mehner2010}).
The interaction of the primary wind with the WBs environment (WBE) is the subject of our study.

{}From HST/STIS spectra of Fe~II, [Fe~III], [Ni~II] and [Ni~III] lines from WBs C and D,
\cite{Smith2004} found the radial velocity of the WBs to be $\sim 40 \km \s^{-1}$, and concluded that the WBs were originated in the LE.
On the other hand, \cite{Dorland2004} presented HST observations of WBs C and D together with simulations of their propagation and
concluded that they formed during the period 1910 -- 1942, preferably during a brightening event which took place in 1941.
For the purpose of this paper it is immaterial when the dense WBE was ejected, but the uncertainty should be kept in mind.

The most controversial property of the binary system is the direction of the semi-major axis in the equatorial plane,
the so called longitude angle $\omega$ (the argument of periapsis).
It is defined such that $\omega = 270^\circ$ for the case where the secondary is closest to us at apastron passages, and it is
furthest away from us during periastron passages.
For $\omega = 90^\circ$ the secondary is furthest from us at apastron passages, while it is
closest to us at periastron passages.
There is no dispute on the orbital plane location.
The disputing camps have a literally opposite view, with one camp arguing for $\omega \simeq 270^\circ$
(e.g., \citealt{Moffat2009}; \citealt{Okazaki2008}; \citealt{Henley2008}; \citealt{Hamaguchi2007};
\citealt{Nielsen2007}; \citealt{Gull2011}; \citealt{Madura2011}, \citeyear{Madura2012}; \citealt{MaduraGroh2012}; \citealt{Parkin2009}),
while the other camp arguing for $\omega \simeq 90^\circ$
(e.g., \citealt{AbrahamGoncalves2007}; \citealt{FalcetaGoncalves2005}; \citealt{KashiSoker2008, KashiSoker2009c}).

The quest for the longitude angle is not a small technical curiosity. Its value is connected to the type of interaction between the winds of the two stars
and between the winds and the CBM.
For example, if the secondary is towards us near periastron passages, $\omega \simeq 90^\circ$, then the
several-weeks decline in the X-ray emission occurring around periastron passages must be accounted for by accretion of the primary wind by the secondary star
\citep{KashiSoker2009a, KashiSoker2009b, Akashi2012}. This is because the dense primary wind cannot account for much X-ray obscuration.  This in turn,
implies that accretion was much more vigorous during the GE, with implications to the energy source of the GE.
Namely, a value of $\omega \simeq 90^\circ$ supports the claim for accretion as the extra energy source of the GE \citep{KashiSoker2010a},
hinting to an accretion event as the energy source for other transient outbursts \citep{KashiSoker2010b, KashiSoker2011}.

In this paper we examine the interaction of the primary wind with a simple model of the WBE.
In section \ref{sec:velocity} we describe the flow structure and previous works.
The numerical setup is described in section  \ref{sec:numerical}.
The interaction of the primary wind with the WBE model is described in section \ref{sec:interaction}, and a comparison between our simulation results and observations is conducted in section \ref{sec:observations}.
Our short summary is in section \ref{sec:summary}.

\section{THE VELOCITY STRUCTURE AT HUNDREDS OF AU}
\label{sec:velocity}

The recent dispute over the value of the longitude angle is centered around the interpretation of the velocity
structure at hundreds of AU from the binary system.
\cite{Gull2009} presented observations of velocity structures of broad ($\sim 500 \km \s^{-1}$) [Fe~II] emission line which extend to
$\sim 1600 \AU$ from the binary system, and [Fe~III], [Ar~III], [Ne~III] and [S~III] lines which extend up to a projected distance of
only $\sim 700 \AU$ from NE to SW.
\cite{Gull2009} observed all those forbidden lines disappearing during the spectroscopic event -- the several weeks period near periastron passages
when large variations in emission and absorption properties occur.
\cite{Madura2012} took HST/STIS spectra of [Fe~III] emission lines from slits in different position angles through the central source
of \etc.
While the low velocity component of these lines was associated with the WBs, \cite{Madura2012} interpreted the high velocity components
of these lines as being formed by winds collision gas as it flows away from the binary system.
\cite{Gull2011} presented more HST/STIS high-ionization forbidden-line observations taken after the 2009 spectroscopic event,
this time in the form of complete intensity maps.

Based on comparison with SPH simulations and under the assumption (which we dispute in the present paper) that the lines
originate by the colliding-winds material as it flows outward, \cite{Gull2009}, \cite{Madura2011}, \cite{Madura2012} and \cite{Gull2011}, concluded that
the high velocity components of the forbidden-line observations fit an orientation in which the secondary
star is in the direction of the observer for most of the binary orbit.
Namely, the longitude angle (the argument of periapsis) is $\omega \simeq 270^\circ$ according to their interpretation.

\cite{Mehner2010} also presented HST/STIS spectra of high ionization lines ([Fe~III] and [N~III])
around $\eta$ Car extending over more than a full cycle.
They also found that most of the emission arrives from the WBs region.
Moreover, they found a variation of the intensity of the lines over the entire orbital cycle,
in addition to the fast increase before periastron that was followed by the typical decrease of the spectroscopic event.
\cite{Mehner2010} identified a fast blue-shifted component of the high ionization lines that appears concentrated near the
stars and elongated perpendicular to the bipolar axis of the system.
\cite{Mehner2010} suggested that this component is related to the equatorial outflow and/or to
dense material known to exist along our line of sight to the system.

\cite{Madura2012} performed 3D numerical simulations of the colliding stellar winds, but completely ignored the slow dense ejecta
in the equatorial plane -- the WBE.
The high resolution observations reported recently by \cite{Gull2011} and \cite{Madura2012} that extend to $\sim 0.^{\prime \prime}5$
from the binary system overlap with the WBE as mapped by \citet{Chesneau2005}.
\cite{SokerKashi2012b} reexamined the observations and interpretation of \cite{Gull2011} and \cite{Madura2012}, and concluded
that the interaction with the WBE cannot be ignored when modeling emission from that region.
Hence, \cite{SokerKashi2012b} claimed, the conclusion of \cite{Madura2012} that $\omega \simeq 90^\circ$ cannot account for the velocity map does not hold.
Instead, \cite{SokerKashi2012b} suggested that the contribution of the winds collision gas to the observed velocity structure at hundreds of AU is small.
The main contribution to the asymmetric flow structure is of the freely expanding primary wind as it collides with the WBE.
Because the WBE is closer to us, and for most of the time the secondary is away from us for $\omega \simeq 90^\circ$, the secondary wind does
not intercept the primary wind flowing towards the WBE.

{{{ Observations by \cite{Smith2003} present evidence of latitude-dependent velocity of the $\eta$ Car primary wind.
\cite{Groh2012}, on the other hand, attribute this observed dependence of the primary wind velocity on the latitude to its collision
with the secondary wind.
In our model, the winds collision occurs on the other side of the binary system,
and has no effect on the WBE region.
Hence, in our simulations we choose a simple radial wind, independent of the latitude. }}}

In the scenario proposed by \cite{SokerKashi2012b} the primary stellar wind collides with the slow dense material
in the WBE and goes through a shock wave.
The post-shocked gas reaches a temperature of $\sim 10^6 \K$ but rapidly cools to $\sim 10^4 \K$ and is compressed.
This warm dense gas is expected to become a source of some emission lines, including forbidden lines.
\cite{SokerKashi2012b} further showed that the intensity and location of the different velocity components of the [Fe~III]~$\lambda$4659
extended forbidden line can be explained by the interaction of the primary wind with the WBE.
\cite{SokerKashi2012b} made a schematic drawing of the flow structure but did not conduct any numerical simulations of the flow.

\begin{figure}[h!]
\begin{center}
\includegraphics[width=0.7\textwidth]{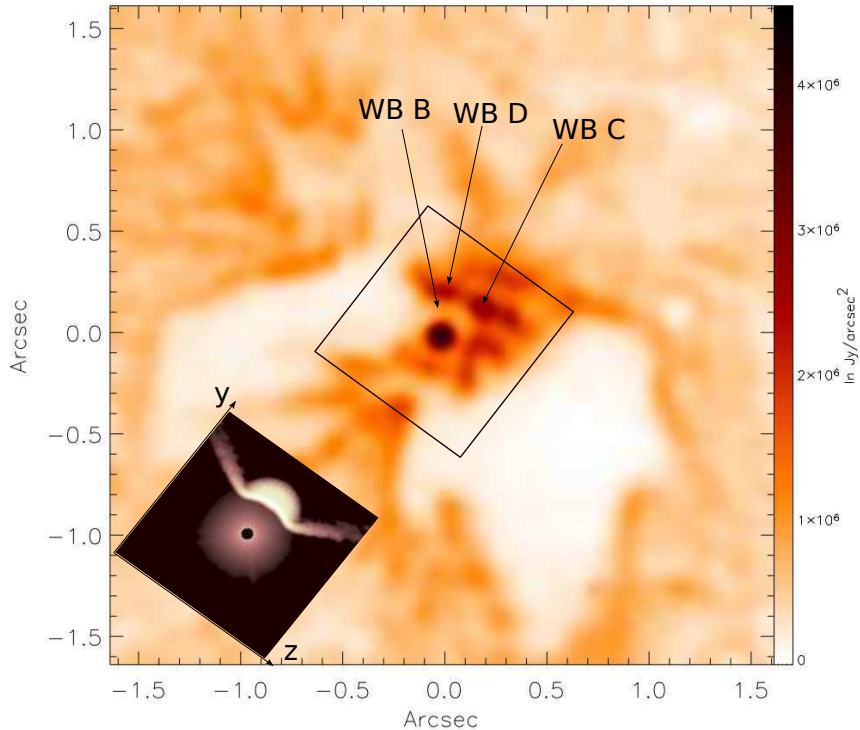}
\caption{The dense Weigelt blobs (WB) and their environment, the WBE, within hundreds of AU from $\eta$ Car,
taken from \cite{Chesneau2005}{}.
The image is a deconvolved image of the emission line Pfund $\gamma$. At the distance of $\eta$ Car each
arcsec is 2300 AU. WBs B, C and D reside in the equatorial plane, inclined to the line of sight by $42^\circ$.
In the inset we show the density {{{in $y-z$ plane}}} of one of our simulations (see Section \ref{sec:interaction}).
The inset corresponds to the square marked around the center of the image.}
\label{fig:WBE}
\end{center}
\end{figure}

In Figure \ref{fig:WBE} we present the observed inner region taken from Figure 5 of \cite{Chesneau2005}.
It is evident that the region surrounding the dense WBs, the WBE, around hundreds of AU on the NW direction (upper right in the image) is quite complicated.
The exact structure is not known as we see material in projection. It is assumed that the dense gas is near the equatorial plane that is inclined to
line of sight by $42^\circ$.
In the present paper we limit ourselves to show that the general velocity structure can be
explained by the interaction of the primary wind with the WBE, as we cannot reproduce accurately the exact density near the equatorial plane.
It will be clear that by rebuilding the WBE in the numerical code we can reproduce the flow structure.
A big uncertainty is the structure of the dense gas and its velocity when it was formed a century ago.
Another uncertainty in reproducing the image is the emissivity at each point due to the lack of knowledge of the ionizing radiation flux from the secondary.
Indeed, the velocity maps change over the orbital period.

Considering the above, we take very simple WBE structures, either one or two spherical dense blobs (clouds).  More details are given in section \ref{sec:numerical}.
\section{NUMERICAL SETUP}
\label{sec:numerical}
The simulations are performed using the high-resolution multidimensional hydrodynamics code FLASH 4.0b \citep{Fryxell2000}.
We employ a full 3D uniform grid (all cells have the same size) with Cartesian $(x,y,z)$ geometry.
We define the numerical $x$ axis to lie along the line of sight.

The numerical $y$ and $z$ axes are defined so that the center of the primary star, that is the source of the wind, the observer and the centers of the clouds that form the WBE in our model are all in the $x-y$ plane ($z = 0$). The WBE model is symmetric about the $z=0$ plane. The $z$ axis and the vector from the primary star to the center of the large WBE cloud define a plane that corresponds to the equatorial plane of the binary system. The grid size is $3000 \AU \times 3000 \AU \times 3000 \AU$ in the $x$,$y$,$z$ directions, with 128 equal-size cubical cells along each axis (single cell size is $\sim 24\AU$). The wind of the $\eta$ Car primary star is modelled by a radial outflow from the central region of the grid (the inner 100 AU), with a mass loss rate of $\dot{M}_{1}=3\times10^{-4}M_{\odot}\yr^{-1}$ and velocity of $v_{1}=500~\rm{km}~{\rm s}^{-1}$ \citep{Pittard2002}.

The WBE has been ejected by the system more than a century ago. However, some dense equatorial outflow might have been continued for tens of years. It is impossible to know the exact mass loss history and mass loss rate that led to the formation of the WBE. Considering that our goals are (a) to show that the interaction of the primary wind with the WBE cannot be neglected, and (b) to show that this interaction can in principle give the velocity maps at hundreds of AU from the system, we consider two very simple models. In those models the WBE is composed either from one or two spherical clouds that start with zero velocity. A more realistic value would be to start the WBE with some radial outflow velocity of $\sim 40 \km \s^{-1}$.
{{{However, if we choose to model the outflow velocity with a non-zero value, we would need to change the
initial conditions of the simulation so that the center of the WBE is closer to the center of the binary system,
and then let the system evolve over 50 years. As there are still many uncertainties regarding the past evolution of
both the WBE and the primary stellar wind, such a simulation would require additional parameters related
to the evolution of the WBE and the primary stellar wind to be added to our model.
Such delicate modelling of the past evolution is not in the focus of the present paper,
and we choose a simplified model of the system taking the radial outflow velocity to be zero.}}}
Another uncertainty is the history of the primary wind, that could have been weaker until recently, or could have been stronger in the equatorial plane until recently. It is hard to know.  We simply base our two models on the present existence of the WBE and its location.
Considering these, we run our simulations for $\sim 50 \yr$,  until the wind launched from the center of the grid fills the entire grid.  That our WBE started with zero velocity rather than $40 \km \s^{-1}$ should be kept in mind while comparing our results with observations.

\begin{figure}[h!]
\begin{center}
\includegraphics[width=0.8\textwidth]{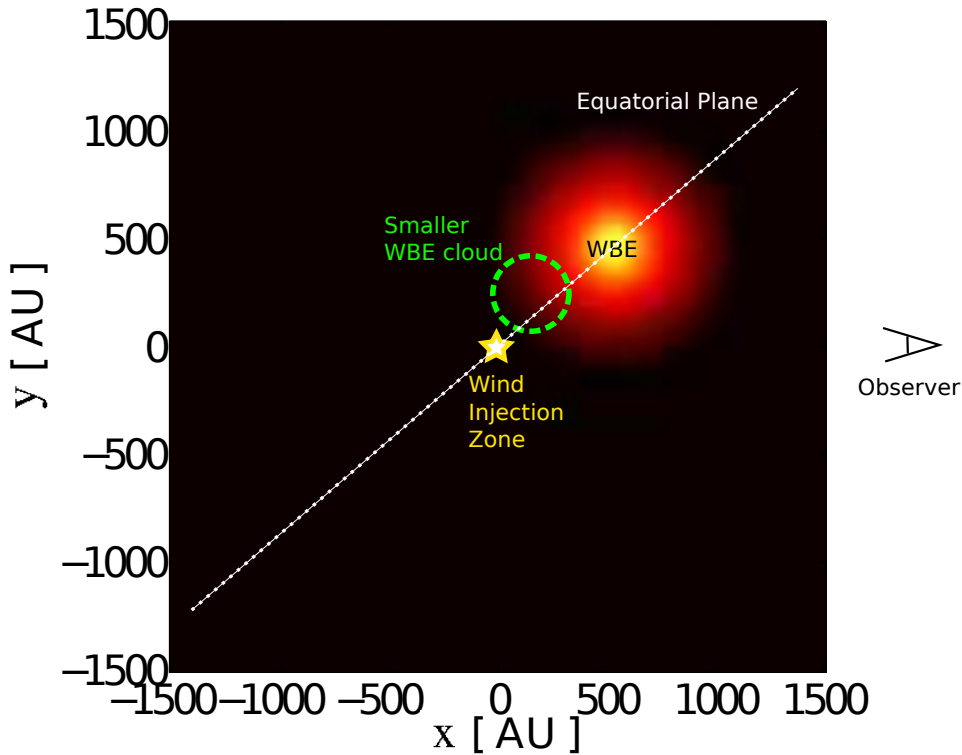}
\caption{The initial setup of the numerical grid for the one-cloud model.
A 2D plane of the 3D computational domain is shown, where the third axis is $-1500 \AU < z < 1500 \AU$.
The location of the second cloud in the two-clouds model is marked by the green dashed circle.
The equatorial plane of the binary system, which is taken to be where most of the WBE mass is,
is defined by the $z$ axis and the line from the the center of the grid to the center of the large WBE cloud.
The intersection of this plane and the $x-y$ plane is marked by the white dotted line.
The observer is in the positive direction of the $x$ axis in the $z=0$ plane. }
\label{fig:Initial}
\end{center}
\end{figure}

The one-cloud structure is composed of a dense spherical cloud centered $700 \AU$ from the center of the primary, as depicted in the inset of Figure \ref{fig:WBE} and in Figure \ref{fig:Initial}. The mass of the cloud is $M_{\rm WBE1}=0.2M_{\odot}$, and the number density in the center of the cloud is $n_{\rm WBE}=10^{9}\cm^{-3}$ ($\rho_{\rm WBE} = 1.03\times10^{-15}$ \g\cmcub). We assume that the cloud extends to a radius of $600 \AU$, and its density profile is given by $\rho_{\rm Cloud1} =  \rho_{\rm WBE} \times (\frac{70 \AU}{70 \AU + r_{\rm WBE1} })$ for $ 0 <  r_{\rm WBE1} < 600 \AU$, so that the initial density at $r_{\rm WBE1} = 600 \AU$ is $\simeq 0.1 \rho_{\rm WBE}$.

In the two-clouds structure a smaller cloud with a radius of $r_{\rm WBE2}=200 \AU$ is added to the one-cloud structure
in the $x-y$ plane, at a distance of $300 \AU$ from the primary star, and at an angle of $30^{\circ}$ from the equatorial plane
(or $72^{\circ}$ from the line of sight, which is the positive $x$ axis in our simulation).
The mass of the smaller cloud is $M_{\rm WBE2}=0.06M_{\odot}$, but we take it to be denser than the large cloud.
The density in the center of the smaller cloud is $2 \times \rho_{\rm WBE}$, and its density profile is given by
$\rho_{\rm Cloud2} =  2 \times \rho_{\rm WBE} \times (\frac{200 \AU}{200 \AU + r_{\rm WBE2}})$ for
$ 0 < r_{\rm WBE2} <200\AU$, so that the initial density at $r_{\rm WBE2} = 200 \AU$ is $\simeq \rho_{\rm WBE}$.
The location and properties of the second cloud are chosen almost arbitrarily.
The sole goal is to demonstrate that small changes in the WBE structure can have noticeable effect on the velocity maps.
As we do not know the exact structure of the WBE, at this stage we limit ourselves to qualitative study.

The initial temperature of the primary wind and the WBE is $T_{i}=1000K$. We add radiative cooling for $T>10^{4}K$,
using the solar composition cooling function $\Lambda(T)$, as given by  \cite{SutherlandDopita1993}.
We impose outflow boundary conditions on all 6 sides of the simulation box.
The initial numerical setup is drawn schematically in Figure \ref{fig:Initial}.
\section{WIND-WBE INTERACTION}
\label{sec:interaction}
\begin{figure}[h!]
\begin{center}
\includegraphics[width=0.8\textwidth]{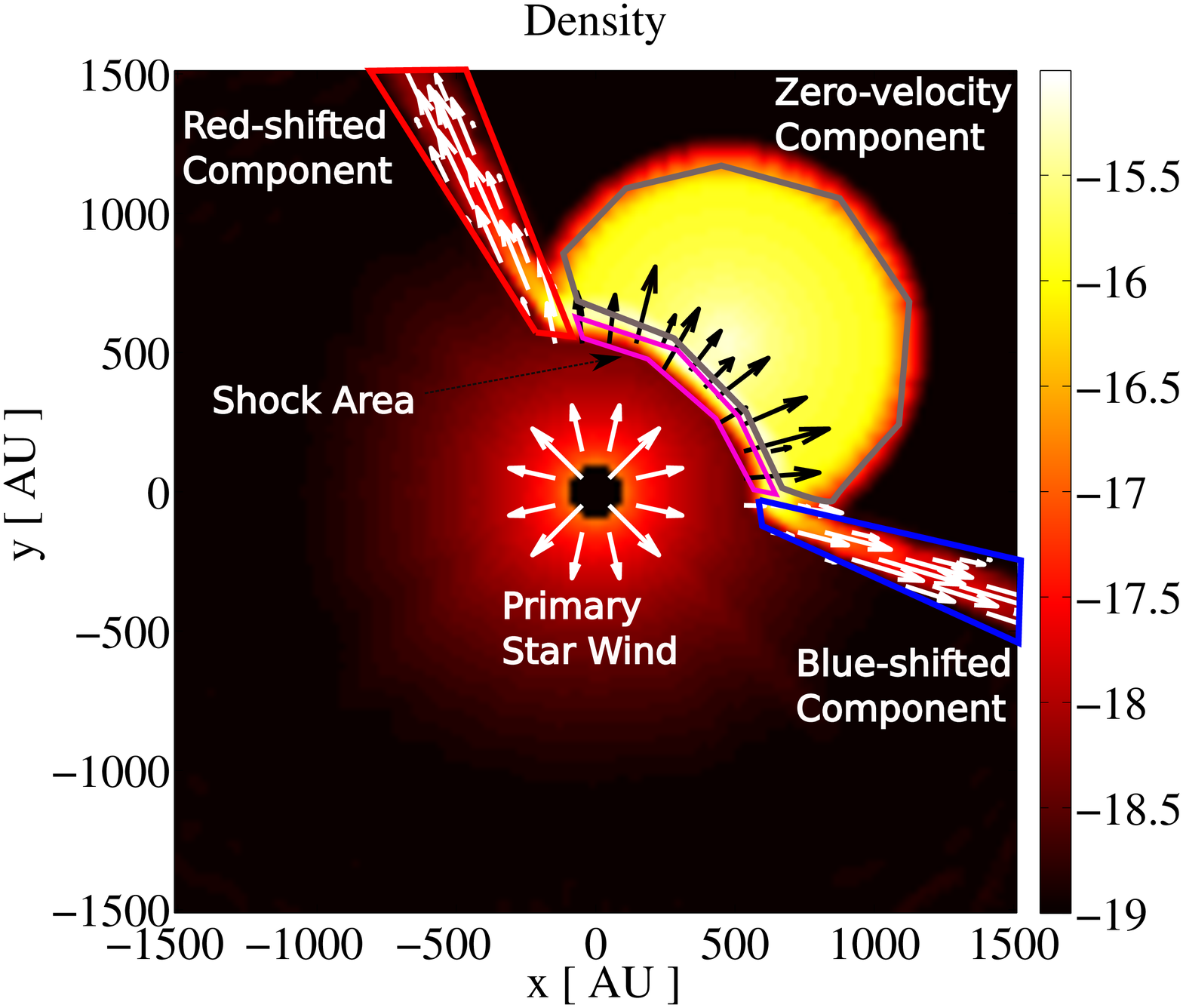}
\caption{The simulated interaction between the $\eta$ Car primary wind and the WBE in the $x-y$ plane. The density scale is logarithmic in units of $\g \cm^{-3}$.  The observer is on the right side of the figure looking left. The WBE is closer to the observer than the central binary system is, and inclined to the line of sight by $42^\circ$. The post-shock material moving towards the observer in the dense-warm region is encircled by a blue line, while the post-shock material moving away from the observed is encircled by a red line. The WBE itself constitutes most of the zero-velocity component. The velocity vectors of the post-shock material and the initial wind are marked by arrows (the different colors of the arrows are for clarity of presentation only). Velocity arrows of the pre-shock primary wind are not plotted, as the flow is a simple radial outflow. The length of the arrows is proportional to the velocity, with a maximum at $500$\kms.}
\label{fig:Interaction}
\end{center}
\end{figure}

We present our results after 50 years of simulation, when the post-shock primary wind reached the boundary of the simulation grid.
In Figure \ref{fig:Interaction} we present the density and velocity maps in the $x-y$ plane of the one-cloud model for the WBE (see Section \ref{sec:numerical} and the inset of Figure \ref{fig:WBE}). Regions that contribute to the red, blue, and zero Doppler shifts are marked.
The primary wind hits the WBE and a shock is formed.
At the apex, where the flow hits the shock wave at a straight angle,
the immediately post-shock temperature is calculated to be $3 \times 10^6 \K$.
The immediate post-shock temperature decreases away from the apex because the shock becomes more oblique.
In any case, the post-shock gas very rapidly {{{(over a timescale of $\sim 6$ months \citep{SokerKashi2012b})}}}
cools radiatively to a temperature of $ T \sim 10^4 \K$ , and
a temperature of $\sim 10^6 \K$ is not seen in the temperature map.
Actually, the post-shock gas {{{moves a distance of  $\sim 10 \AU$, less than a cell size}}}, before it cools to $T \sim 10^4 \K$.
A warm and dense gas now occupies the post-shock regions.
It is this gas that is the main source of the radiation observed in the velocity maps according to our model.
As can be seen in Figure \ref{fig:Interaction},
the freely flowing primary wind is diverted from its initial radial direction as it passes through the shock,
and it flows around the cloud. In our interpretation of the velocity map \citep{SokerKashi2012b}
the post-shock gas flowing away from (towards) us forms the red-shifted (blue-shifted) component.
Because of the inclination by $42^\circ$ of the equatorial plane (the line from the center of
the grid to the large WBE cloud in our model) to our line of sight the blue and red-shifted components are not symmetric,
as we show in the next section.
{{{ In addition to that, the shocked material (both blue and red-shifted components)
is further accelerated in the post-shock area. This is because of two processes that
take place:
(a) The high pressure in the post-shock region accelerates the wind that flows around the WBE.
(b) The primary wind hits the shocked region at a large distance from the blob at a very small angle,
producing a highly oblique shock. The post-shock gas near the outer surface of the post-shock outflow
is re-accelerated by this newly shocked primary wind.}}}

\begin{figure}[htp!]
\begin{center}
\includegraphics[width=0.8\textwidth]{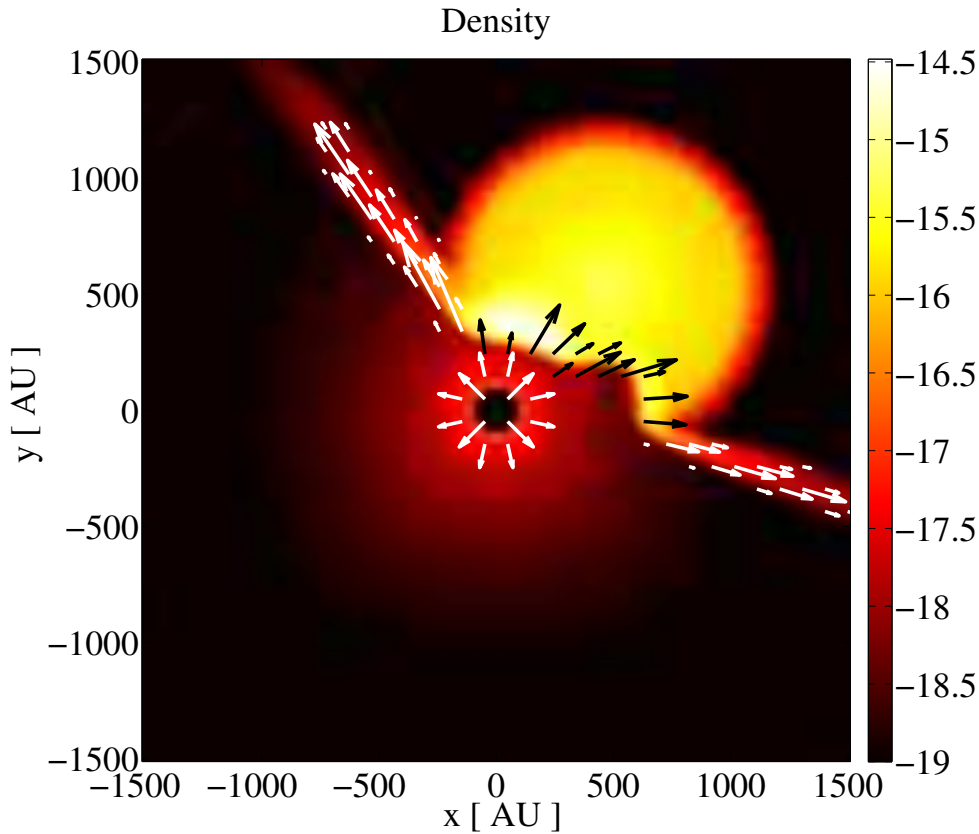}
\caption{Same as Figure \ref{fig:Interaction}, but for the two-clouds model. The smaller cloud is in the bottom left part of the large cloud.}
\label{fig:Density2}
\end{center}
\end{figure}

For comparison with observations later on (Section \ref{sec:observations}) we build another very simple model for the structure of the WBE. In the second model for the WBE a smaller cloud is added to the cloud in the one-cloud model, as shown schematically in Figure \ref{fig:Initial}. The smaller cloud is denser than the large cloud, and is located closer to the primary star (for more details see Section \ref{sec:numerical}). The resulting density and velocity map for the two-clouds model is shown in Figure \ref{fig:Density2}.

\begin{figure}[htp!]
\begin{center}
\includegraphics[width=0.48\textwidth]{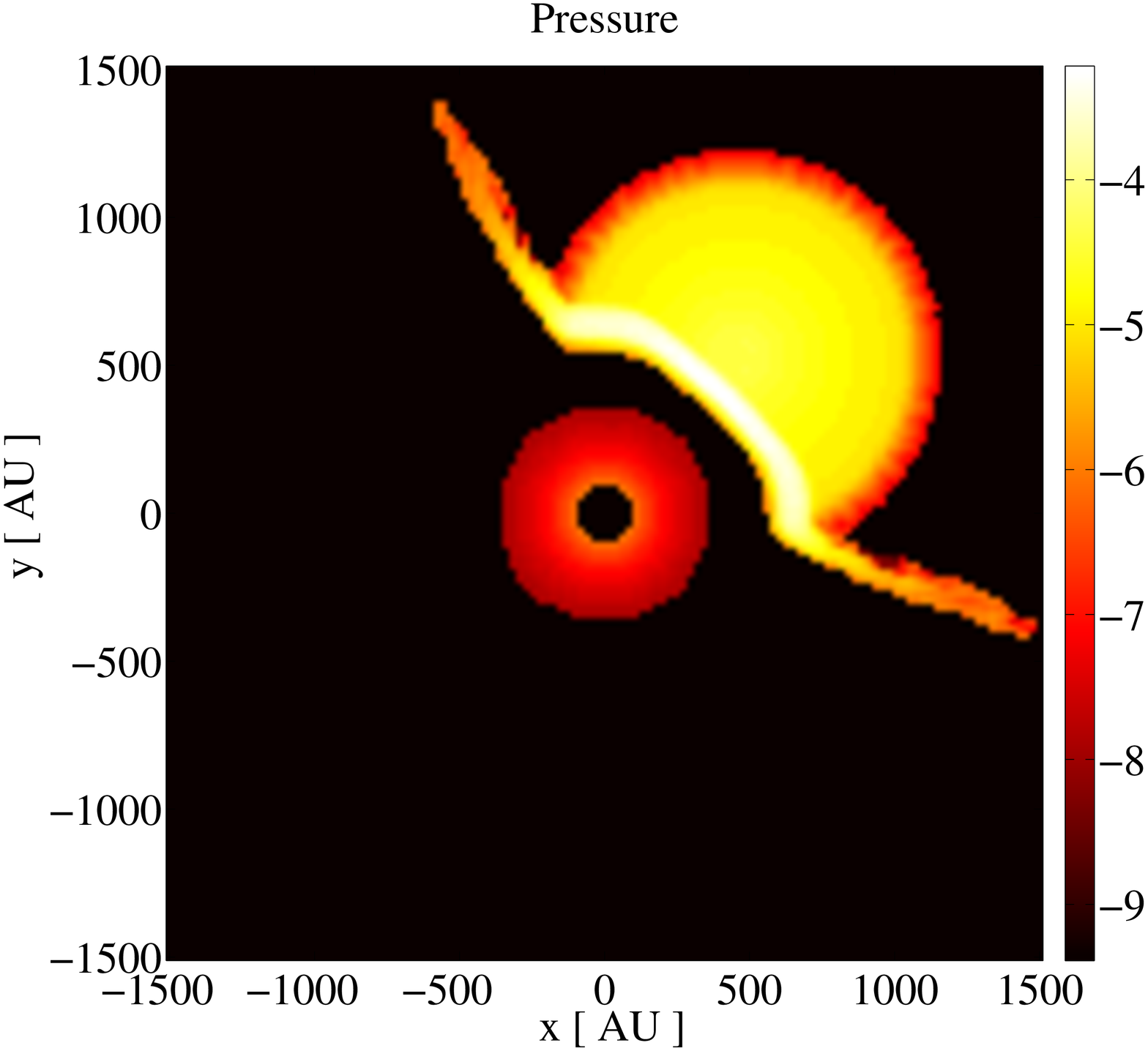}
\includegraphics[width=0.47\textwidth]{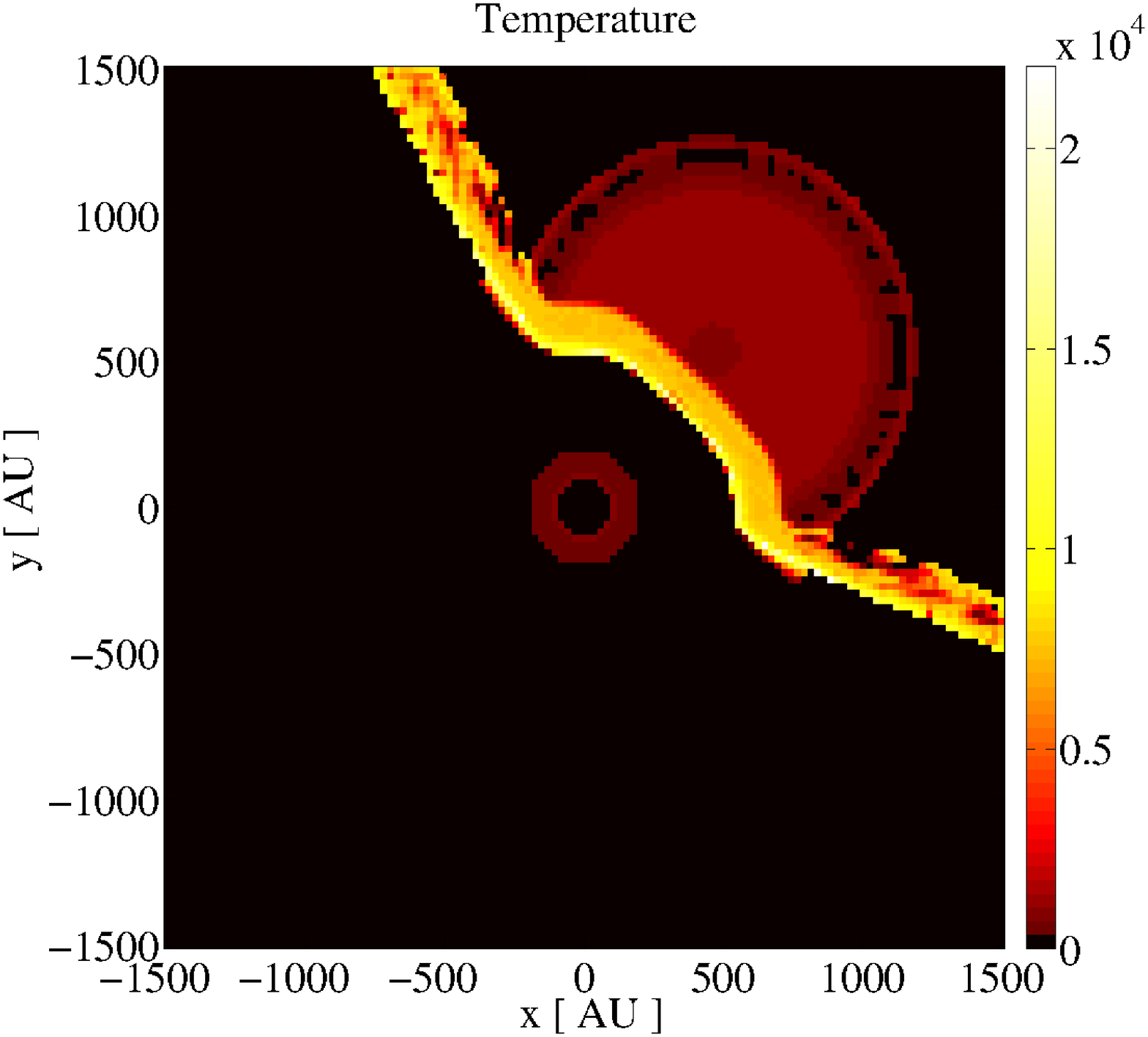}\\
\includegraphics[width=0.48\textwidth]{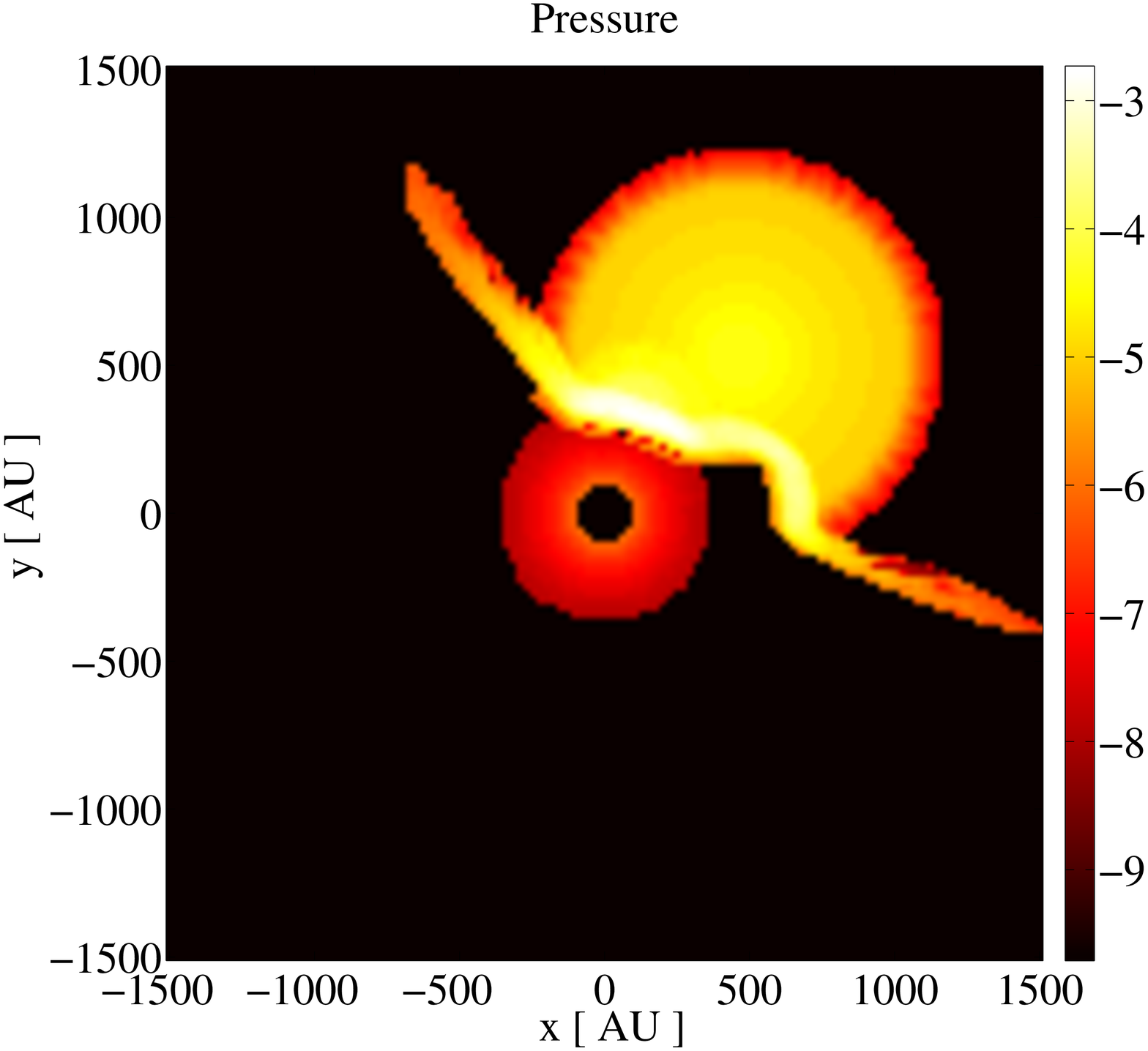}
\includegraphics[width=0.47\textwidth]{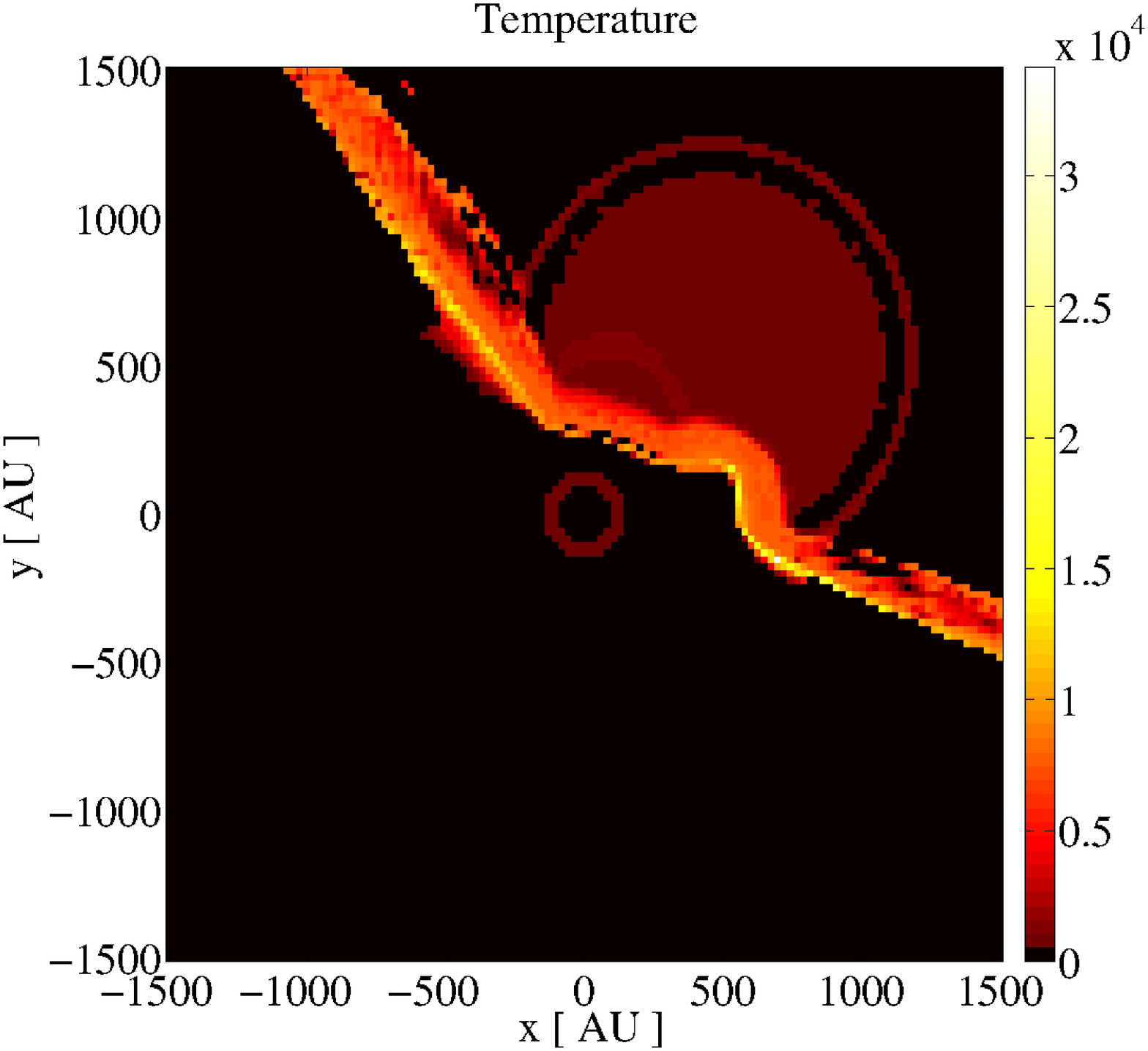}
\caption{The pressure (left column) and temperature (right column) maps for the one-cloud (top row) and two-clouds (bottom row) models. The scale in the pressure maps is logarithmic in units of   $\rm{dyn}\cm^{-2}$. The temperature maps are in$\K$. }
\label{fig:PressureTemperature}
\end{center}
\end{figure}

The pressure and temperature maps of both models are shown in Figure \ref{fig:PressureTemperature}. It is evident from the temperature maps that the post-shock primary wind cools very rapidly
\citep{SokerKashi2012b}. As the iron responsible for the [Fe~III] line is ionized, only the warm regions ($T \ga 7500 \K$) contribute to the emission.
However, it should be kept in mind that we do not include any radiative heating or ionization by the stellar radiation, and so we cannot quantify the contributions of the different warm regions.
{{{ The WBE itself is cold relatively to the shocked area. The outskirts of the WBE expand at about the
speed of sound in the WBE, which is $\sim 2 $\kms{}. Over 50 years, the WBE expands by $\sim 20\AU$.
As the initial radius of the WBE, both in the one-cloud and the two-clouds models, is $\sim 600 \AU$,
this expansion has little effect on the simulations. }}}
 The pressure map shows the high pressure of the clouds that divert the primary wind.
\section{COMPARISON WITH OBSERVATIONS}
\label{sec:observations}

In comparing the velocity maps of our models to observations we are not only limited by the lack of knowledge of the mass distribution in the WBE, but also by the fact that the ionization stage of iron and the ionizing radiation from the secondary star are poorly determined \citep{Madura2012, SokerKashi2012b}.
The variation of the velocity maps with orbital phase \citep{Madura2012, Gull2011} testify to the complexity of the problem.
As our goal is to demonstrate the general velocity maps expected from the interaction of the primary wind with the WBE,
we use a very simple prescription. We assume that only gas with temperatures of $T > 7500 \K$ contributes to the [Fe~III]
line and that the the emissivity goes as the density square ($\rho^2$).
This very simple approach should be kept in mind when examining our comparison with observations.
For example, as we do not consider the ionization radiation from the secondary star, we have no dependence on orbital phase.
Because of these uncertainties in our modeling, we basically make a qualitative comparison.

To obtain the velocity components distribution maps, we integrate $\rho^{2}$ along the line of sight for three velocity bins. The velocity bins are based on the bins in \cite{Gull2011} ($-400$ to $-200$ \kms, $-90$ to $30$\kms{} and $+100$ to $+200$\kms), but we extend the boundaries of the three velocity bins by $30$\kms{} to each side, to account for the random
velocity of gas in the post-shock regions.

\begin{figure}[htp!]
\begin{center}
\includegraphics[width=0.55\textwidth]{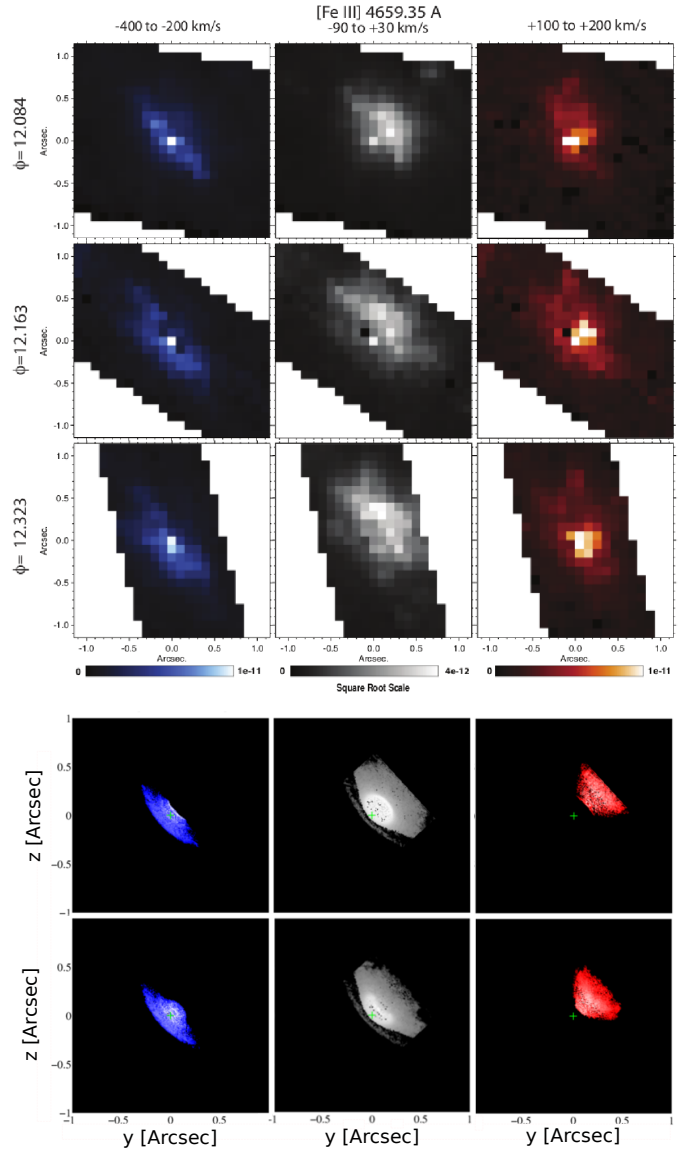}
\caption{Top 3 rows show the observed velocity components at three orbital phases \citep{Gull2011}, with intensity in square root scale. The two bottom rows are synthetic maps of our simulations, with intensity in logarithmic scale. The fourth row from the top shows the velocity maps obtained for the single-cloud WBE model. The bottom row shows the velocity maps for the two-clouds WBE model (see Section \ref{sec:numerical}). The velocity bins in the simulated maps are extended to each side by $30 \km \s^{-1}$ to take into account random motion of the post-shock gas.}
\label{fig:Gull}
\end{center}
\end{figure}

We compare our results with the observations of \cite{Gull2011} in Figure \ref{fig:Gull}. The observations at three orbital phases are shown in the first three rows (taken from Figure 2 of \citealt{Gull2011}). Orbital phase $12.0$ is the spectroscopic event of January 2009, and the orbital period is 5.53 years \citep{Damineli2008}. The fourth and fifth rows of Figure \ref{fig:Gull} show the synthetic velocity maps of our one-cloud and two-clouds models, respectively. The intensity in the observed maps (top three rows) is in square root scale, while the intensity in the synthetic maps (bottom two rows) is in logarithmic scale.
As we do not follow the ionization degree of iron and the ionizing radiation, we cannot compare the absolute intensity values in our simulations to the observed values. Hence, we limit the discussion to a qualitative comparison between the spatial morphologies of the observed and the simulated maps.

\begin{figure}[htp!]
\begin{center}
\hspace{-1 cm}
\includegraphics[width=0.35\textwidth]{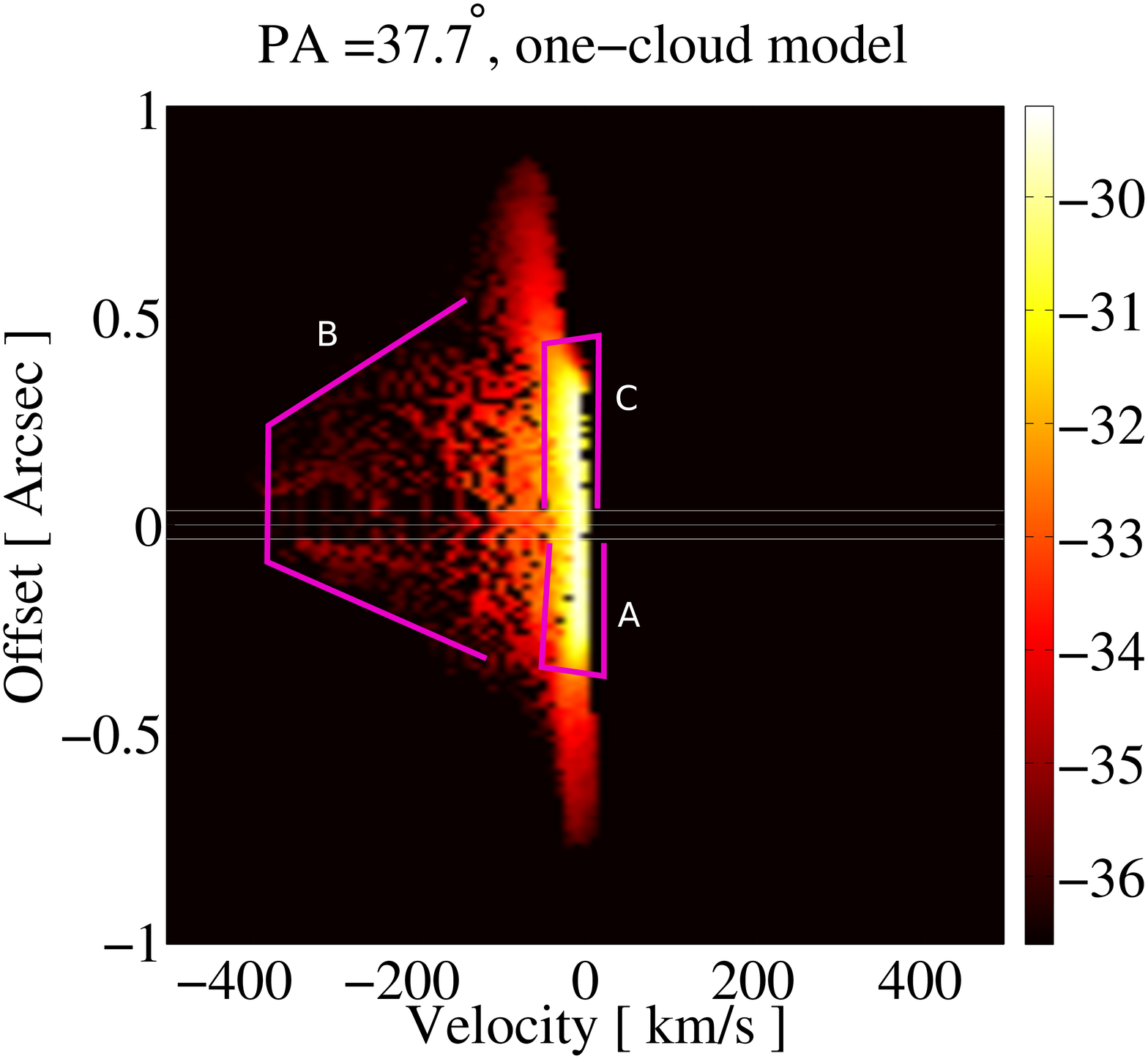}
\includegraphics[width=0.35\textwidth]{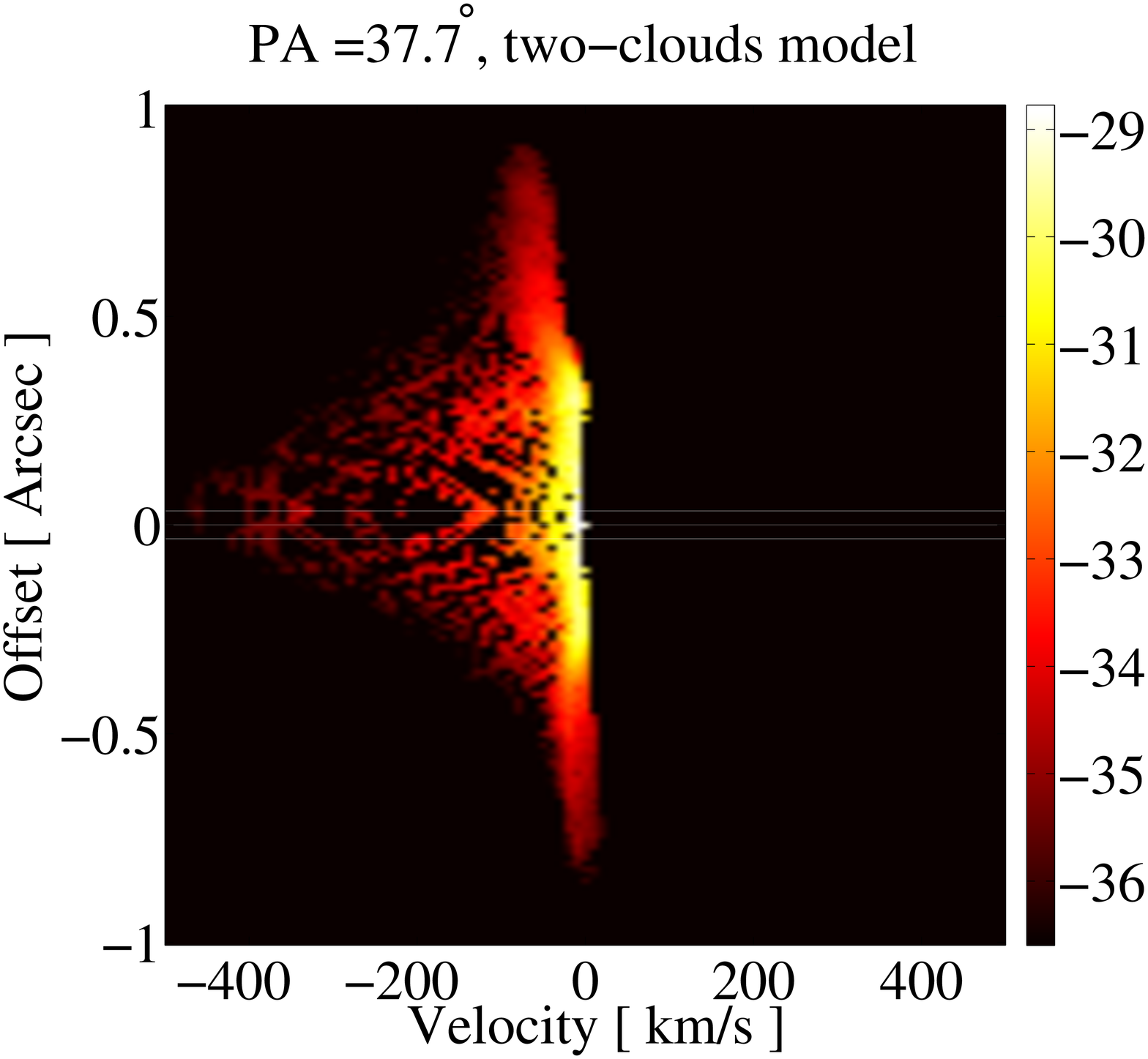}
\includegraphics[width=0.33\textwidth]{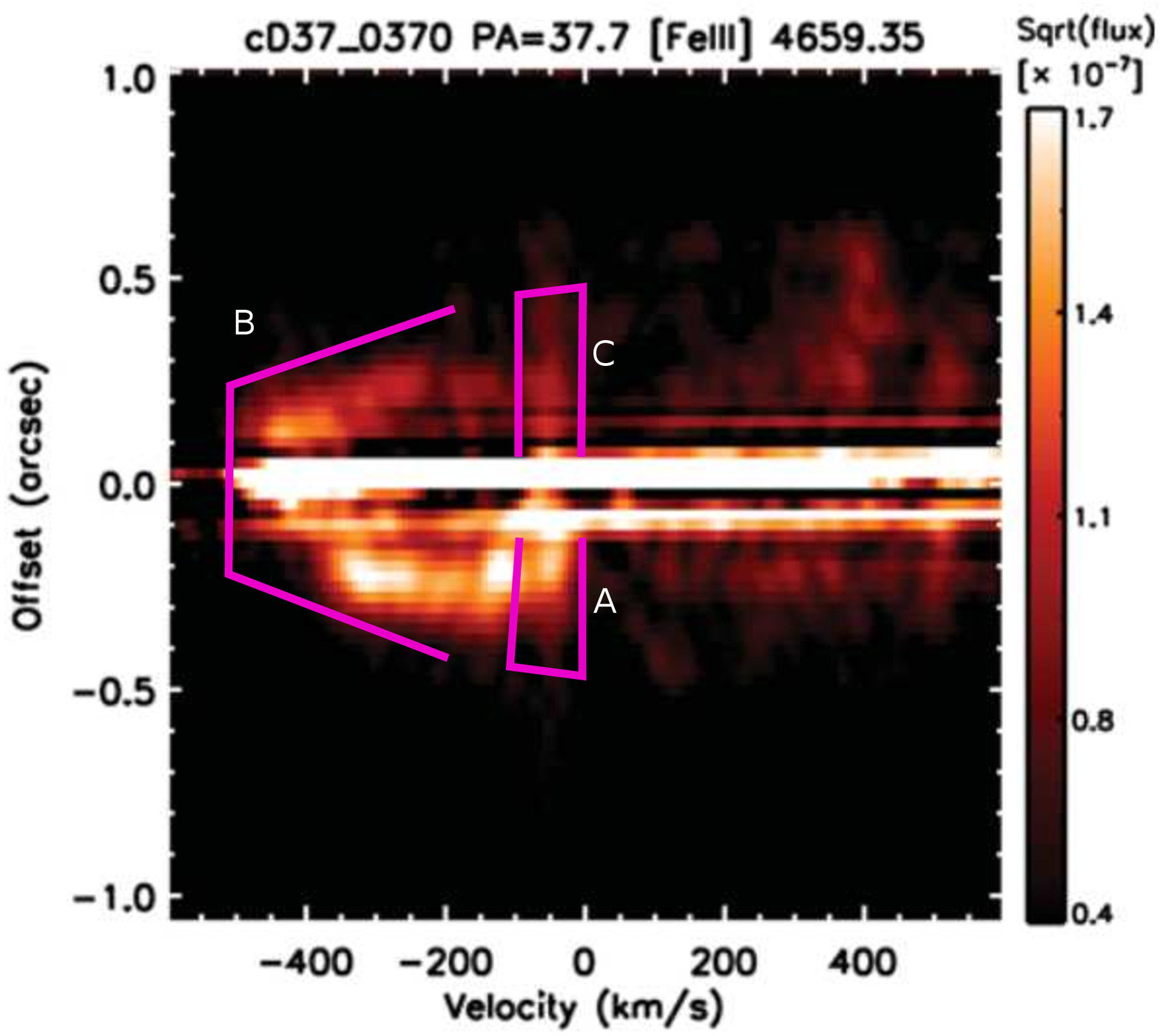}
\caption{A qualitative comparison of the velocity along position angle ${\rm  PA} = +37.7^\circ$.
The right panel is the observed velocity profile in the [Fe~III] line taken from \citep{Madura2012}.
The color code in the synthetic maps is of the intensity calculated by integrating over $\rho^2$ along line of sight.
Units are logarithmic and in relative values.
Our synthetic profiles are given for the one-cloud and two-clouds models in the left and middle panels,
respectively. As we do not calculate the ionization degree of iron and the ionizing radiation,
the comparison is qualitative and limited to identifying large scale features.
Features A and C mark low velocity components. {{{Note that these features should appear at a velocity of $\sim -40 \km \s^{-1}$,
but appear at $0$ \kms{} due to our simplifying choice of zero velocity for the WBE. }}} Feature B marks the blue-shifted velocity component.
The middle panel shows the same line profile for the two-clouds model.
Although the two models seem to be almost similar, the small differences demonstrate that
with correct construction of the WBE the velocity profile can be changed to better match observations.
The emission from the center of the binary system (within $0.^{\prime \prime} 05$) is not
 modelled in our simulations, and is covered by a white line.}
\label{fig:Slit38}
\end{center}
\end{figure}

\begin{figure}[htp!]
\begin{center}
\hspace{-1 cm}
\includegraphics[width=0.35\textwidth]{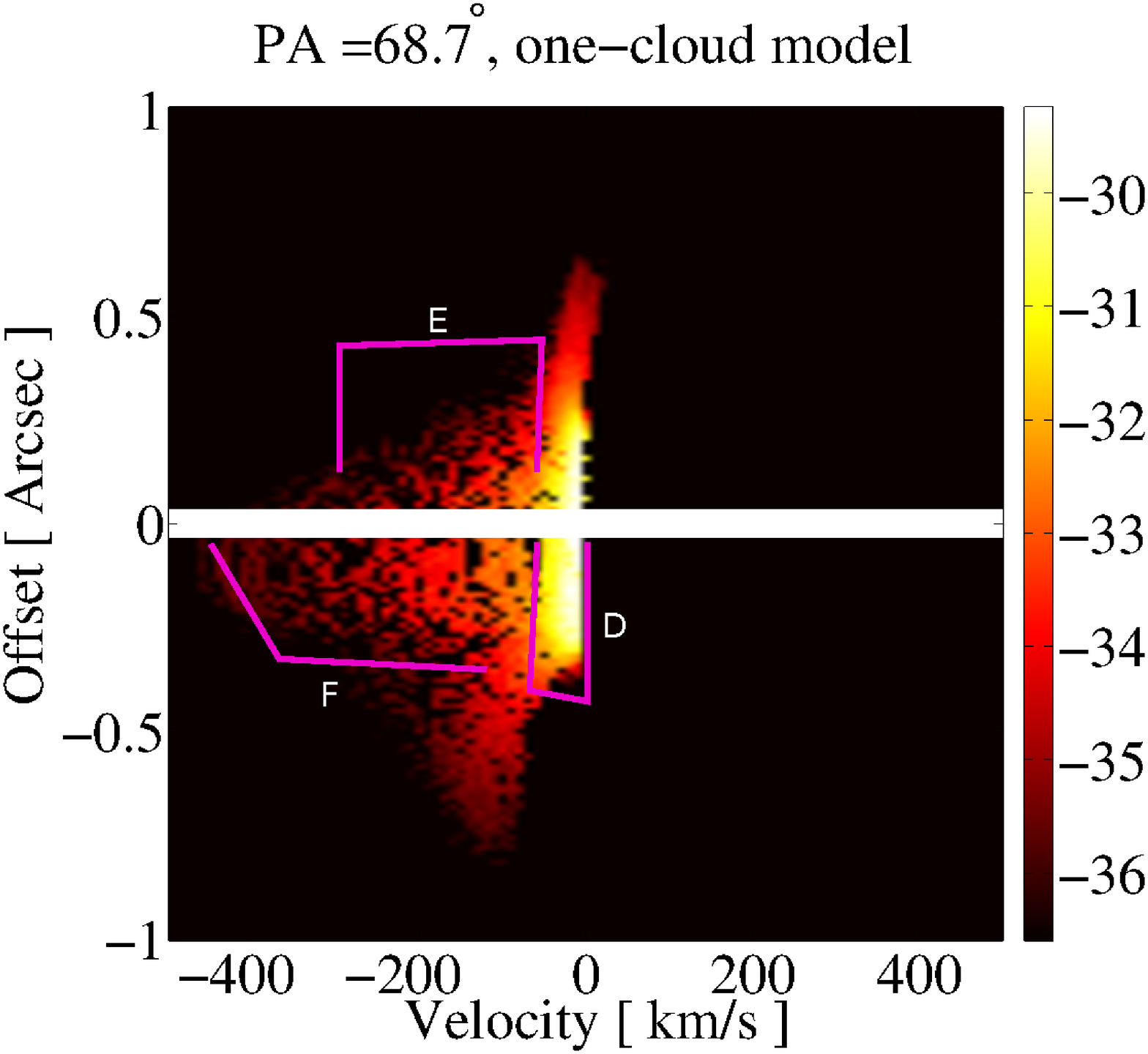}
\includegraphics[width=0.35\textwidth]{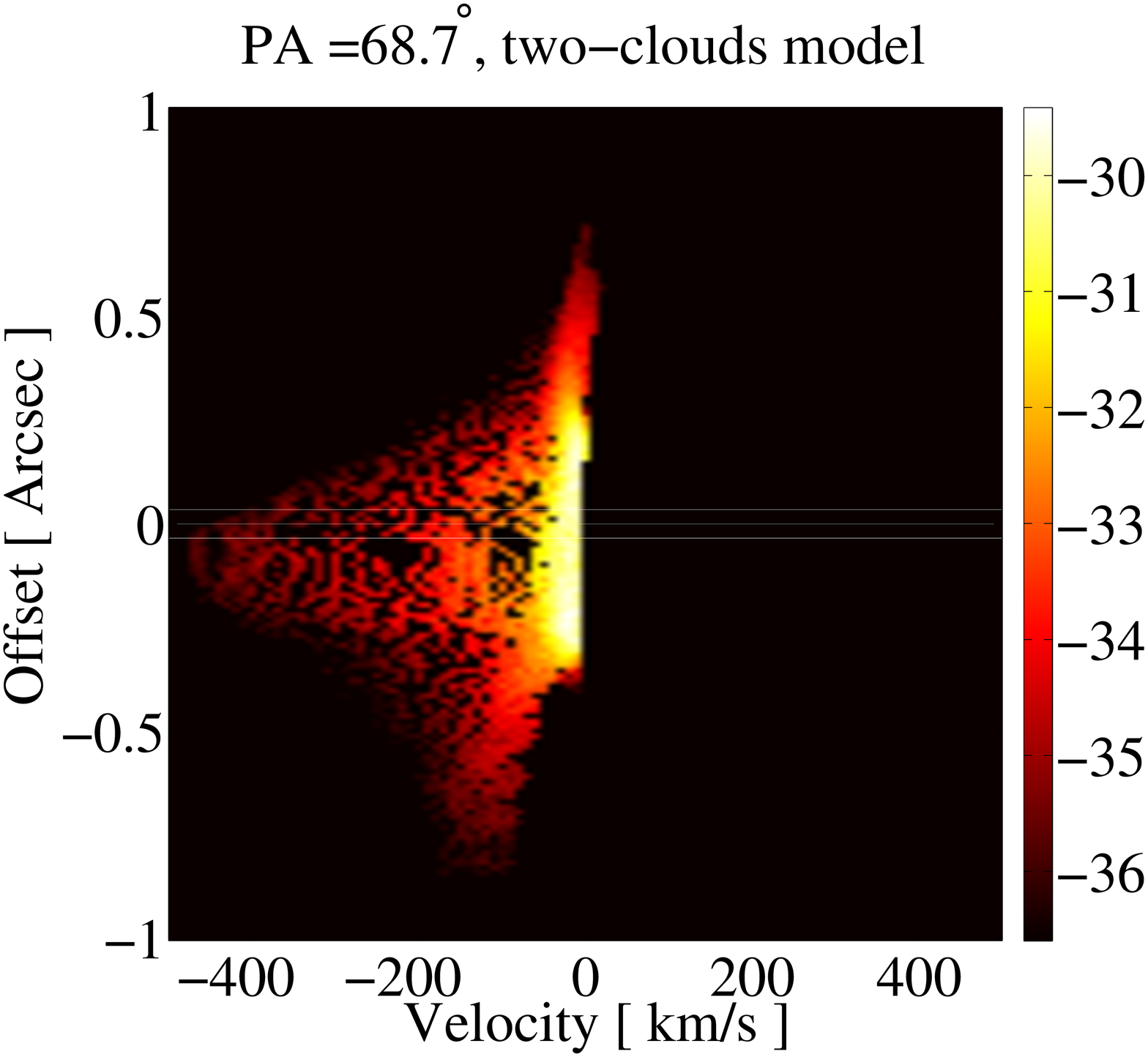}
\includegraphics[width=0.33\textwidth]{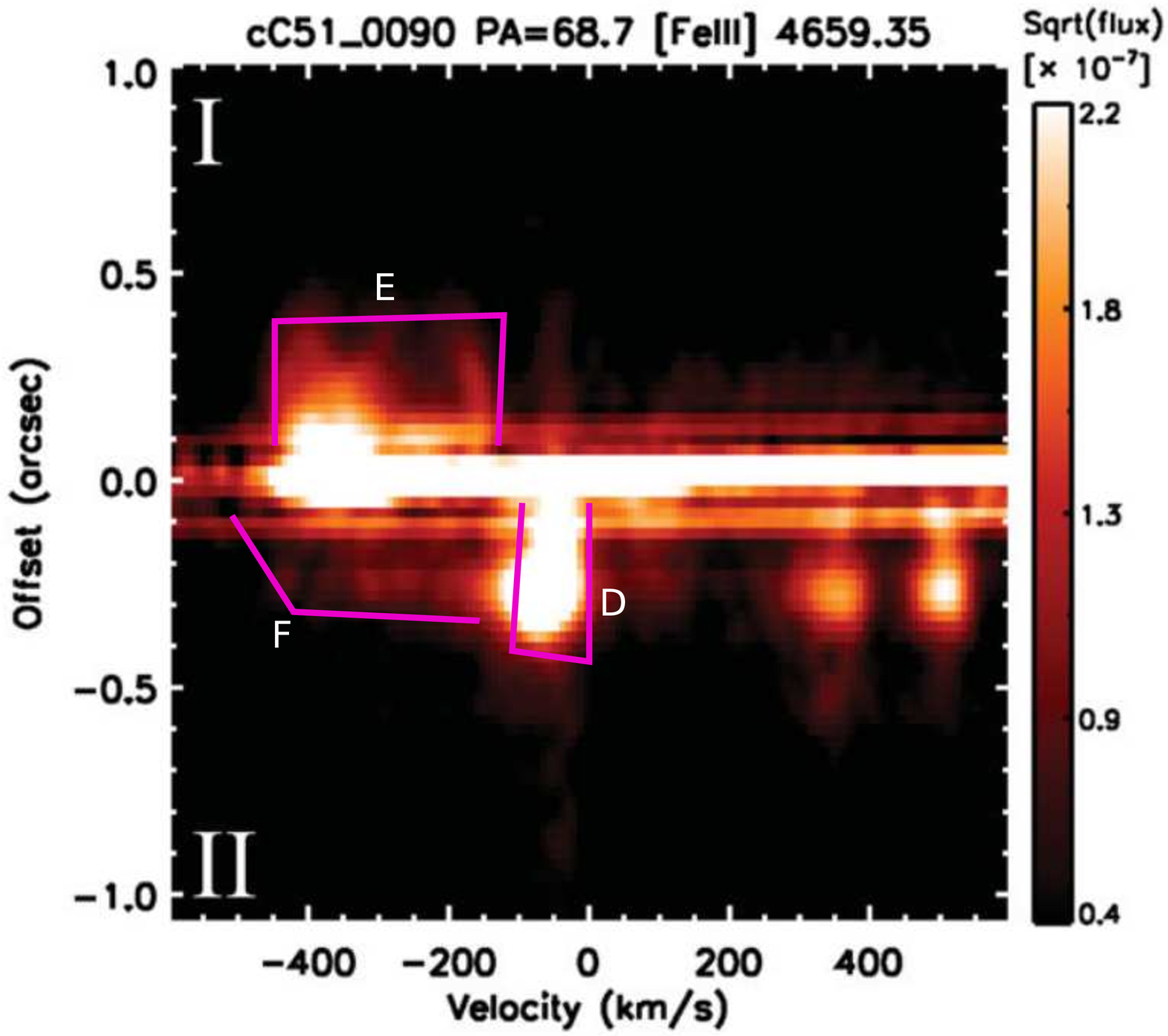}
\caption{ Same as Figure \ref{fig:Slit38}, but for  ${\rm  PA} = +68.7^\circ$.
Feature D marks the low velocity component. Features E and F mark blue-shifted velocity components.}
\label{fig:Slit69}
\end{center}
\end{figure}

The overall morphology, both in the observed and simulated blue-shifted components (left column) is along the NE-SW direction, crossing the center but with some extension to the SE (lower left in the figure). The zero-velocity component (middle column) results mostly from the warm ($T > 7500 \K$) WBE and slow wind around it, and reproduces quite well the qualitative structure of the observed velocity map. We see that even our simplified choice of initially spherical WBE in the one-cloud model manages to qualitatively reproduce the observed zero-velocity map. The red-shifted component (right column) in our simulations recreates the asymmetrical red-shifted component that appears in the NW region of the observed velocity map.
The qualitative agreement with observations of the red component is better in the two-clouds model,
as can be seen in the bottom-right panel of Figure \ref{fig:Gull}.
The small differences between the one-cloud and two-clouds models demonstrate that by different
reconstructions of the WBE we can change the synthetic velocity maps.
In principle we can find very good matching with observations.
This is beyond the goals of the present paper as the parameters space of the WBE modeling is huge.
Overall, we consider the qualitative agreement to be satisfactory considering
the mentioned uncertainties and unknowns of our modeling procedure.
We also compare the results of our simulations with the HST/STIS spatially resolved [Fe~III] spectra of $\eta$ Car,
as given by \cite{Madura2012}. We reproduce the spectro-images for various position angles (PA)
of the HST/STIS relative to the direction of the North in $\eta$ Car observations. The comparison is presented in Figures \ref{fig:Slit38} and \ref{fig:Slit69}, and,
as before, is only qualitative in nature.

The agreement between the synthetic results and the observed line profiles for PA=$+37.7^\circ$ (see Figure \ref{fig:Slit38}) is poor.
However, considering the unknowns discussed above we cannot expect a good match.
None the less, we can identify observed features in our synthetic maps, as marked by letters in
Figures \ref{fig:Slit38} and \ref{fig:Slit69}. The blue-shifted velocities from $0$\kms{} to $-400$\kms{}
are marked as feature B in Figure \ref{fig:Slit38}.
The low-velocity components are marked as features A and C in Figure \ref{fig:Slit38}).
For PA = $+68.7^\circ$ (see Figure \ref{fig:Slit69}), we reproduce the low velocity component
present in the observed line profile at $-40$ \kms{} (marked as feature D in Figure \ref{fig:Slit69}).
We recall that in our simulations the velocity of the simple WBE model is $0$ \kms{},
corresponding to the velocity value of feature D in the synthetic line profile.
A more detailed and realistic construction of the WBE would give the WBE a velocity of $\sim -40 \km \s^{-1}$.
There is some agreement between the general features of the observed and the
synthetic line profiles for blue-shifted velocities (marked as features E and F in Figure \ref{fig:Slit69}).

\newpage
\section{SUMMARY}
\label{sec:summary}
We perform 3D gas flow simulations of the interaction of the $\eta$ Car primary wind with the dense gas clouds (WBE) in the equatorial plane of the binary system. Considering the complex and vaguely resolved structure of the WBE (see Figure \ref{fig:WBE}), we choose two simple models for the WBE: (1) The WBE is modelled by a single spherical cloud; (2) The WBE is modelled by two spherical clouds, one large and one small (see Figure \ref{fig:Initial} for more details). For comparison with [Fe~III] observations \citep{Gull2011} we construct synthetic emission maps at different velocity intervals (see Figure \ref{fig:Gull}). The exact reconstruction of the synthetic maps is made difficult by the high uncertainties in the ionization degree of iron in the WBE, and in the varying amount of ionizing radiation produced by the secondary star. Taking these into account we use a simple prescription to calculate the emission. We assume that the emission is produced only by warm gas with temperatures of $T > 7500 \K$, and take the emission intensity to be proportional to $\rho^{2}$. This set of simplifying assumptions allows only qualitative comparisons. As can be seen in Figure \ref{fig:Gull}, we can qualitatively explain the velocity maps in the red, zero, and  blue-shifted velocity intervals. The observed zero-velocity component is explained by the presence of slow dense warm material that constitutes the shocked wind and the WBE.

We also compare the results of our simulations with the HST/STIS spatially resolved [Fe~III] spectra of $\eta$ Car, as given by \cite{Madura2012}. While we are able to reproduce some of the large scale features in the observed line profiles, the overall agreement between the observed and the synthetic line profiles is somewhat limited. To obtain better agreement the WBE structure must be better resolved, and the ionization degree of iron and the ionizing radiation must be calculated. These are complicated tasks that are much beyond the present study.

Based on the results of our simulations we conclude that the observed emission maps and line profiles can be explained as a result from the interaction between the primary star wind and the WBE, as previously suggested by \citep{SokerKashi2012b}. As the WBE is closer to the observer than the center of the binary system is, it follows from our results that the primary star (which blows the freely expanding wind) is also closer to the observer than the secondary star for the majority of the binary period, e.g. the longitude angle of the binary system is $\omega\simeq90^\circ$. Our results add to a growing number of observations in favour of this longitude angle, and dispute the claim made by \cite{Gull2011} and \cite{Madura2012} that the observed velocity maps and line profiles exclude a value of $\omega \simeq 90^\circ$. Our result shows that this orientation can account for the observations, and that the collision of the primary wind with the WBE cannot be ignored.  We admit though that the winds collision should also be considered. Including both winds collision and the WBE interaction is very demanding numerically, and should be the goal of future studies.

Knowledge of the exact orientation of the $\eta$ Car binary system is of importance for correct interpretation of observations in the upcoming 2014 spectroscopic event. Knowing the correct value of $\omega$ will allow us to better understand the interaction of the binary system near periastron passages, and aid in further study of the accretion processes that took place during the Great and Lesser Eruptions, deepening our understanding of interacting LBV binary systems.

{{{We thank the referee, Ricardo Gonzalez, for helpful comments.}}} We thank Amit Kashi for useful comments.
This research was supported by the Asher Fund for Space Research at the Technion, and the USA-Israel Binational Science Foundation.
The FLASH code used in this work is developed in part by the US Department of Energy under Grant No. B523820 to the Center for Astrophysical Thermonuclear Flashes at the University of Chicago.
The simulations were performed on the TAMNUN HPC cluster at the Technion.

\end{document}